\documentclass[a4paper,11pt]{article}
\usepackage{jheppub} % for details on the use of the package, please
\usepackage{amsthm} %for dealing with theorem environments,

\usepackage{stmaryrd}

\title{Gauge/gravity duality and lattice simulations of one dimensional SYM with sixteen supercharges}

\author[a]{Daisuke Kadoh}
\author[b]{and Syo Kamata}

\affiliation[a]{KEK Theory Center, High Energy Accelerator Research Organization (KEK),\\
            \ Tsukuba, Ibaraki 305-0801, Japan}
\affiliation[b]{Department of Physics, Rikkyo University, \\ Tokyo 171-8501, Japan}
\emailAdd{kadoh@post.kek.jp}
\emailAdd{skamata@rikkyo.ac.jp}

\preprint{
{\flushright
\vspace{-8mm}
KEK-TH-1759, RUP-14-11\\
}}

\abstract
{
We study the gauge/gravity duality for supersymmetric $SU(N)$ Yang-Mills theory in 1+0 dimension 
with sixteen supercharges using lattice simulations.
The conjectured duality states that the gravity side is described by $N$ D0-branes in type IIA superstring at large $N$, and   
the thermal gauge theory reproduces the black hole thermodynamics at low temperature.
In this paper, we explain the Sugino lattice action used in the simulations in detail, and examine the supersymmetric
 Ward-Takahashi identity to confirm the restoration of supersymmetry in the continuum limit. 
We also estimate the internal energy of the black hole from the lattice results for $N=14,\,32$, and find that it
smoothly approaches the prediction of the gravity side as the temperature decreases.  
}


\begin{document}

\maketitle
%\flushbottom

%%%%%%%%%%%%%%%%%%%%%%%%%%%%%%%%%%%%%%%%%%%%%%%%%%%%%%%%%%%%%%%%%%%%%%%%%%%%%%%%%%%%%%%%%%%%%%%%%%%%
%                                                                                                                                                   %
%                                                                                                                                                   %
%               Introduction                                                                                                                     %
%                                                                                                                                                   %
%                                                                                                                                                   %
%%%%%%%%%%%%%%%%%%%%%%%%%%%%%%%%%%%%%%%%%%%%%%%%%%%%%%%%%%%%%%%%%%%%%%%%%%%%%%%%%%%%%%%%%%%%%%%%%%%%

\section{Introduction}
\label{sec:intro}

The gauge/gravity duality is an equivalence between  
strongly coupled gauge theory and the theory of gravity on a curved space, which is regarded as
a generalization of the original AdS/CFT correspondence \cite{Maldacena:1997re}. 
At present, it is widely accepted, and many related studies have been carried out in various fields of physics \cite{Klebanov:2000me}.
However, it is not a theorem but a conjecture. 
In the verification of the duality conjecture, numerical simulations of the lattice gauge theories will play a key role.
In this paper, we perform lattice simulations of one dimensional supersymmetric  $SU(N)$  Yang Mills theory
with sixteen supercharges ("the BFSS model" \cite{Banks:1996vh}) and examine the duality in this system.

The gravity dual of this theory is described by $N$ D0-branes in type IIA
superstring in the large $N$ limit \cite{Itzhaki:1998dd}, and the behavior of the thermal gauge theory at low temperature %($T/\lambda^{1/3} \ll 1$)
reproduces that of a black hole \cite{Klebanov:1996un}. However, at low temperature, the gauge theory is a  strongly coupled one 
in which known analytic techniques such as the perturbation theory become less effective.
In addition, localization methods \cite{Pestun:2007rz} and other analytic treatments \cite{Erickson:2000af}
 for BPS states are not applicable to the present case because supersymmetry is broken by temperature.
Therefore, some numerical approaches are needed to examine the duality.

Numerical simulations of this gauge theory have been carried out independently by two collaborations: 
M.~Hanada, J.~Nishimura and others \cite{Anagnostopoulos:2007fw}, \cite{Hanada:2008ez}, \cite{Hanada:2009ne}, \cite{Hanada:2013rga}  and 
S.~Catterall and  T.~Wiseman \cite{Catterall:2007fp}, \cite{Catterall:2008yz}, \cite{Catterall:2009xn}. 
\footnote{ Lattice simulations of two dimensional SYM with sixteen supercharges have been performed
in \cite{Catterall:2010ya}, \cite{Giguere:2015cga}.
}
Some regularization schemes are needed to put the theory on a computer.
Hanada et al$.$ used the momentum sharp cutoff, 
while Catterall-Wiseman employed the standard lattice theory.
In both the theories, all of the supersymmetries are broken not only by temperature but also by the regularization schemes.
However, as they expected, there would be no serious problems because the SUSY breaking effects of the regularization schemes disappear in the continuum limit thanks to the UV finiteness of one dimensional gauge theory.
Some obtained results imply the validity of the duality. 
Nevertheless, further simulations are required to verify it with high accuracy.

In this paper, we aim to verify the duality conjecture from lattice simulations of the gauge theory side.
In particular, we employ the Sugino lattice action with keeping two supercharges on the lattice \cite{Sugino:2003yb}, \cite{Sugino:2004qd}, \cite{Sugino:2004uv}.
In section 2, we briefly review the continuum theory and its twisted action, after which 
the lattice action is defined from the twisted one. 
Section 3 describes some algorithms and techniques used in the simulations. 
We show numerical results of the supersymmetric Ward-Takahashi identity in section 4, then present the internal energy of the black hole in section 5.
Finally, section 6 is devoted to a summary.\\
%We give the definitions of SYM with sixteen superchrges in several dimensions in appendix \ref{sym_16susy}.
%The full lattice action and the corresponding lattice Dirac operator are given in appendix \ref{full_lattice_action}.

%%%%%%%%%%%%%%%%%%%%%%%%%%%%%%%%%%%%%%%%%%%%%%%%%%%%%%%%%%%%%%%%%%%%%%%%%%%%%%%%%%%%%%%%%%%%%%%%%%%%
%                                                                                                                                                   %
%                                                                                                                                                   %
%              One dimensional SYM with 16 supercharges                                                                             %
%                                                                                                                                                   %
%                                                                                                                                                   %
%%%%%%%%%%%%%%%%%%%%%%%%%%%%%%%%%%%%%%%%%%%%%%%%%%%%%%%%%%%%%%%%%%%%%%%%%%%%%%%%%%%%%%%%%%%%%%%%%%%%

\section{One dimensional SYM with 16 supercharges}

In this section, we introduce one dimensional supersymmetric Yang-Mills theory with sixteen supercharges (the BFSS model \cite{Banks:1996vh}).
Using two nilpotent supercharges $Q_\pm$, 
the continuum action is expressed as a $Q_\pm$-exact form, and 
the Sugino lattice action is defined from the exact one \cite{Sugino:2004uv}. 
\footnote{
In this paper, the gauge group is $SU(N)$  and, unless otherwise noted,
all of the fields are expressed as the matrix-valued functions.
The gauge group generators $T^a(a=1,\cdots,N^2-1)$ are traceless hermitian matrices which satisfy the Lie algebra,  $[T^a,T^b]=if_{abc}T^c$,
with the normalization ${\rm tr}(T^aT^b)=\delta_{ab}$,
The scalar and fermi fields, as well as the gauge field $A_0$, are expanded as $\varphi(t)=\sum_{a=1}^{N^2-1}\varphi^a(t) T^a$.
Moreover, in this section, we impose the periodic boundary condition on all of the fields to study the SUSY invariance, while in the following sections the fermions satisfy anti-periodic boundary condition to study the theory at finite temperature.
}

\subsection{Continuum theory}

% Continuum action
Supersymmetric Yang-Mills theory in 1+0 dimension with sixteen supercharges contains a gauge field $A_0$,  nine scalars $X_i \ (i=1,\cdots,9)$
and sixteen fermions $\psi_\alpha \ (\alpha=1,\cdots,16)$.
The continuum euclidean action is given by
\begin{eqnarray}
S_{cont.}
  & =& \  \frac{N}{\lambda} \int {\rm d}t \;  \mathrm{tr}\left\{
        \frac{1}{2} (D_0 X_{i})^{2} 
      - \frac{1}{4} [X_{i},X_{j}]^2 
        \right. \nonumber \\
 &&    \hspace{3cm} \left. 
      + \frac{1}{2} \psi_\alpha D_0 \psi_\alpha 
      + \frac{1}{2}  \psi_\alpha  (\gamma_{i})_{\alpha\beta} [X_{i}, \psi_\beta]  
        \right\},
\label{def:1dSYM_action} 
\end{eqnarray}
where $\lambda$ is the 't Hooft coupling constant and
$\gamma_i$ are real symmetric matrices satisfying the nine dimensional Clifford algebra,
$\{\gamma_i,\gamma_j\}=2\delta_{ij}$.
Also, the covariant derivative $D_0$ is defined 
as $D_0 \ \cdot = \partial_t \  \cdot +i[A_0,\ \cdot \ ]$.%  for adjoint fields $\varphi$. 

% Gauge fields 
In one dimension, the gauge field has no kinetic terms and appears only
through the covariant derivative.
However, as well as four dimensional gauge theory,
it leads to an invariance of the theory
under the gauge transformations,
\begin{eqnarray}
\delta_\omega A_0 =-D_0 \omega,
\quad \delta_\omega \varphi = -i [\varphi,\omega],  
\label{def:1d_gauge_tr}
\end{eqnarray}
for ${}^\forall  \varphi \in\{X_i,\psi_\alpha\}$, and $\omega$ is an infinitesimal gauge parameter.
In this case,  one can fix the gauge completely: $A_0=0$ for no boundary conditions, 
or  $A_0=w$ for periodic time  where $w$ is a constant corresponding to the Polyakov line.  
Nevertheless, we define the theory without any gauge fixing to treat both SUSY and gauge invariance manifestly.

The supersymmetry transformations are given by
\begin{eqnarray}
&&  \delta_\xi A_0    =  \xi_\alpha \psi_\alpha,
\nonumber \\
&&  \delta_\xi X_i  =  -i \xi_\alpha (\gamma_i)_{\alpha\beta} \psi_\beta,
\label{def:1d_susy}
\\
&&  \delta_\xi \psi_\alpha = i (\gamma_i)_{\alpha\beta} \xi_\beta D_0 X_i 
                         - \frac{1}{2} \left(\Sigma_{ij}\right)_{\alpha\beta} \xi_\beta [X_i,X_j],
\qquad \Sigma_{ij} = \frac{i}{2}[\gamma_i, \gamma_j],
\nonumber
\end{eqnarray}
where  $\xi_\alpha$ are sixteen fermionic parameters. 
We can show that the action (\ref{def:1dSYM_action}) is invariant 
under the transformations (\ref{def:1d_susy}) 
by using the Fierz identity for the fermions
and the Jacobi identity for the scalars.
The well-known Noether's theorem tells us that 
the corresponding supercharges $Q_\alpha$ 
are the generators of the infinitesimal transformations, $\delta_\xi=\xi_\alpha Q_\alpha$.

This one dimensional theory can be obtained by a dimensional reduction of four dimensional ${\cal N}=4$ SYM, 
or its topologically twisted theory \cite{Dijkgraaf:1996tz}.
In the topological field theory, Lorentz group is twisted by the internal group of ${\cal N}=4$ SYM.
The transformation laws of the fields and the supercharges under the twisted Lorentz group are different from those 
under the original one.
The two linear combinations of the sixteen supercharges $Q_\pm$ are scalars which are nilpotent up to the gauge transformations.
Furthermore, one can express the action with a topological term as a $Q_\pm$-exact form.
The $Q_\pm$-invariance of the action follows from the nilpotency of $Q_\pm$.
%\footnote{ In appendix A, we give a full account of the target one dimensional theory,
%four dimensional ${\cal N}=4$ SYM and its topologically twisted theory 
%and the relations among them and ten-dimensional ${\cal N}=1$ SYM.  
%}

In order to define the $Q_\pm$-exact action directly from (\ref{def:1dSYM_action}),
we only have to change the field variables to those of the twisted theory.
%
%As twisted varables, 
We use
$A_\mu, B_i, C, \phi_{\pm}$ and $\eta_{\pm}, \psi_{\pm\mu}, \chi_{\pm i}$ 
as the bosonic and  fermionic twisted fields, respectively, for $\mu=0,1,2,3$ and $i=0,1,2$.
Under the four dimensional twisted Lorentz transformations, 
 $A_\mu, \psi_{\pm\mu}$ transform as vectors,
$C,\phi_\pm,\eta_{\pm}$ remain unchanged, (namely are scalars),
%transform as scalars 
and the other bosons and fermions transform as self-dual tensors.
$B_i,\chi_{\pm i}$ are three independent components of the bosonic and fermionic tensors,
 respectively. In one dimension, the indices $\mu,i$ should be regarded as the names of the fields.

The bosonic twisted fields $A_\mu,B_i,C, \phi_{\pm}$ are given by
\begin{eqnarray}
&& X_\mu = A_\mu, \qquad (\mu=1,2,3)
\nonumber
\\
&& X_4 = -B_2, 
\nonumber
\\
&& X_5 =  B_1, 
\nonumber
\\
&& X_6 = -B_0, 
 \label{def:X_to_twisted_boson}
\\
&& X_7 = \frac{1}{2} C, 
\nonumber
\\
&& X_8 = \frac{1}{2} \left( \phi_+ - \phi_- \right),
\nonumber
\\
&& X_9 = \frac{i}{2} \left( \phi_+ + \phi_- \right), 
\nonumber
\end{eqnarray}
with $A_0$ unchanged. 
$A_\mu, B_i,C$ are hermitian and $(\phi_+)^\dag = -\phi_-$.

The fermionic twisted variables $\eta_{\pm}, \psi_{\pm\mu}, \chi_{\pm i}$ and the original variables $\psi_\alpha$ are related to each other by
\begin{eqnarray}
&&
\left( \hspace{-1mm}
  \begin{array}{c}
     \psi_1 \\
     \psi_2 \\
     \psi_3 \\
     \psi_4 \vspace{-4mm}\\
                   \\
     \psi_5 \\
     \psi_6 \\
     \psi_7 \\
     \psi_8 \vspace{-4mm} \\
                   \\       
     \psi_{9} \\
     \psi_{10} \\
     \psi_{11} \\
     \psi_{12} \vspace{-4mm}\\
                   \\
     \psi_{13} \\
     \psi_{14} \\
     \psi_{15} \\
     \psi_{16} \\
  \end{array}
\hspace{-1mm} \right)
= \frac{1}{\sqrt{2}}
\left( \hspace{-1mm}
  \begin{array}{c}
     \psi_{-0}   + \frac{i}{2} \eta_{+} \\
     \psi_{-1}   - i           \chi_{+2} \\
     \psi_{-2}   + i           \chi_{+1} \\
     \psi_{-3}   - i           \chi_{+0} \vspace{-4mm} \\
     \\
     \psi_{+0}   + \frac{i}{2} \eta_{-} \\
     \psi_{+1}   + i           \chi_{-2} \\
     \psi_{+2}   - i           \chi_{-1} \\
     \psi_{+3}   + i           \chi_{-0} \vspace{-4mm} \\
     \\
     -i(\psi_{-0}   - \frac{i}{2} \eta_{+} ) \\
     -i(\psi_{-1}   + i           \chi_{+2}) \\
     -i(\psi_{-2}   - i           \chi_{+1}) \\
     -i(\psi_{-3}   + i           \chi_{+0}) \vspace{-2mm} \\
     \\
     -i(\psi_{+0}   - \frac{i}{2} \eta_{-}) \\
     -i(\psi_{+1}   - i           \chi_{-2}) \\
     -i(\psi_{+2}   + i           \chi_{-1}) \\
     -i(\psi_{+3}   - i           \chi_{-0}) 
  \end{array}
\hspace{-1mm} \right),
\label{def:psi_to_twisted_fermion}
\end{eqnarray}
with the appropriate gamma matrices.
%
%  nine dimensional gamma matrices
% 
\footnote{ We take the following gamma matrices, 
\begin{eqnarray}
&&                 \gamma_1 = \sigma_2  \otimes  \, {\mathbf 1} \,  \otimes  \sigma_3          \otimes  \sigma_2,         
\quad \quad    \gamma_4 = \sigma_3  \otimes  \sigma_2           \otimes  \, {\mathbf 1} \, \otimes  \sigma_2,  
\quad \quad    \gamma_7 = \sigma_3  \otimes  \sigma_1           \otimes  \, {\mathbf 1} \, \otimes  \, {\mathbf 1}, \,
\nonumber \\
&&                 \gamma_2 = \sigma_2  \otimes  \, {\mathbf 1} \,  \otimes  \sigma_2          \otimes  \, {\mathbf 1}, \,\,   
\quad \quad    \gamma_5 = \sigma_3  \otimes  \sigma_2           \otimes  \sigma_2          \otimes  \sigma_3, 
\quad \quad    \gamma_8 = \sigma_3  \otimes  \sigma_3           \otimes  \, {\mathbf 1} \, \otimes  \, {\mathbf 1}, \, 
\label{def:nine-gamma-matrices}
\\
&&                  \gamma_3 = \sigma_2  \otimes  \, {\mathbf 1} \,  \otimes  \sigma_1          \otimes  \sigma_2,   
\quad \quad     \gamma_6 = \sigma_3  \otimes  \sigma_2           \otimes  \sigma_2          \otimes  \sigma_1,  
\quad \quad     \gamma_9 = \sigma_1  \otimes  \, {\mathbf 1} \,  \otimes  \, {\mathbf 1} \, \otimes  \, {\mathbf 1},
\nonumber
\end{eqnarray}
where $\sigma_i$ are the Pauli matrices. 
The gamma matrices act on the fermions 
as 
\begin{eqnarray}
(\gamma_1)_{\alpha_1\alpha_2} \psi_{\alpha_2}=(\sigma_2)_{i_1i_2}  
\otimes {\mathbf 1}_{j_1j_2} \otimes  (\sigma_3)_{k_1k_2} \otimes (\sigma_2)_{l_1l_2} 
\psi_{\alpha_2}
\end{eqnarray}
for $\alpha=8(i-1) +4(j-1)+2(k-1) +l$.}

Let us define two supercharges $Q_\pm$ as follows:
\begin{eqnarray}
&& Q_+ = \frac{1}{\sqrt{2}}(Q_5 + i Q_{13}),\\
&& Q_- = \frac{1}{\sqrt{2}}(Q_1 + i Q_{9}).
\end{eqnarray}
%which correspond to $\eta_{\mp}$. 
From ({\ref{def:1d_susy}}),
we can read $Q_\pm$-transformation laws, 
\begin{eqnarray}
\begin{array}{ll}
   Q_{\pm} A_\mu = \psi_{\pm \mu},
&  \hspace{1.3cm}
   Q_{\pm} \psi_{\pm \mu} = -i D_\mu \phi_{\pm},
 \\ 
   Q_{\pm} B_i   = \chi_{\pm i}, 
&  \hspace{1.3cm} 
   Q_{\pm} \chi_{\pm i} = [B_i, \phi_{\pm}], 
\\
   Q_{\pm} C     = \eta_{\pm},
&  \hspace{1.3cm}  
   Q_{\pm} \eta_{\pm} = [C, \phi_{\pm}],
\\
   Q_{\pm} \phi_{\mp} = \eta_{\mp},
&  \hspace{1.3cm}  
   Q_{\pm} \eta_{\mp} = [\phi_\mp, \phi_{\pm}], 
\label{cont_Q}
\\
   Q_{\pm} \phi_{\pm} = 0,
&
\\
&  \hspace{1.3cm}  
   Q_{\pm} \chi_{\mp i} = \frac{1}{2}[C,B_i] \pm H_i,
\\
&  \hspace{1.3cm}  
    Q_{\pm} \psi_{\mp \mu} = \frac{i}{2} D_\mu C  \pm \tilde H_\mu, 
\vspace{-2mm}\\
\end{array}
\end{eqnarray}
\begin{eqnarray}
\begin{array}{l}
   Q_{\pm} H_i = \pm(              [ \chi_{\mp i}, \phi_\pm ]
                     + \frac{1}{2} [ \chi_{\pm i}, C        ]
                     + \frac{1}{2} [ B_i,      \eta_\pm ] ), \hspace{3mm}
\nonumber \\
   Q_{\pm} \tilde H_\mu =  \pm \left(              [ \psi_{\mp \mu}, \phi_\pm ]
                               + \frac{1}{2} [ \psi_{\pm \mu}, C        ]
                               - \frac{i}{2} D_\mu \eta_\pm \right),
\nonumber \\
\\
\end{array}
\end{eqnarray}
%
%for $\mu=0,1,2,3$ and $i=0,1,2$.
%
where $D_0$ is the covariant derivative and
$D_\mu(\mu=1,2,3)$ are the commutators $D_\mu \cdot =i[A_\mu,\cdot]$. 
Note that $D_\mu$ are used for $\mu=0,1,2,3$ because the twisted theory is originally defined in four dimensions.
Moreover, seven auxiliary fields $H_i$ and $\tilde H_\mu$ are introduced to make the transformations satisfy
\begin{eqnarray}
Q_{\pm}^{2}= i \delta_{\phi_\pm}, \quad \{ Q_{+},Q_{-} \} = -i \delta_{C},
\label{eq:q2}
\end{eqnarray}
where the right-hand sides are the  gauge transformations (\ref{def:1d_gauge_tr}) with the field-dependent gauge parameters.
\footnote{The fields $\phi^a_\pm$ are imaginary, in general.
An imaginary function $\phi$ can be expressed as $\phi=\phi_R + i \phi_I$  where $\phi_R$ and $\phi_I$ are real functions.
Similarly, the gauge transformation $\delta_{\phi}$ means $\delta_{\phi} \equiv \delta_{\phi_R} + i\delta_{\phi_I}$.
}

%
%  Q-exact form
%
Performing the change of variables (\ref{def:X_to_twisted_boson}) and (\ref{def:psi_to_twisted_fermion}) with the gamma matrices (\ref{def:nine-gamma-matrices}),
\footnote{
In addition, we added the following Gaussian integrals to the action,
\begin{eqnarray}
&&
S \rightarrow S + \frac{N}{\lambda} \int {\rm d} t \;  \mathrm{tr}\left\{ 
   \left(
     H_i-i \left( F_{i3} + \frac{1}{2} \epsilon_{ijk}F_{jk} \right) -\frac{1}{2} \epsilon_{ijk} [B_j,B_k] 
   \right)^2
\right.
\nonumber 
\\
&& \hspace{2cm} 
\left.
   + \left(
       \tilde H_i-i(\epsilon_{ijk} D_jB_k + D_3 B_i )
     \right)^2
   + \left(
       \tilde H_3+iD_i B_i 
   \right)^2
\right\}.
\label{eq:theta_term}
\end{eqnarray}
}
the action (\ref{def:1dSYM_action}) can be expressed as the $Q_{\pm}$-exact form,
\begin{eqnarray}
S_{cont.} 
& = &
  Q_{+} Q_{-} \frac{N}{2 \lambda} \int {\rm d}t \; {\rm tr} 
  \left\{
            -2i B_{i} \left( F_{i3} + \frac{1}{2} \epsilon_{ijk}F_{jk} \right)
            -\frac{1}{3} \epsilon_{ijk} B_{i}[B_{j},B_{k}] 
  \right. 
  \nonumber
\\
&& \hspace{3.8cm}
\left.
  - \psi_{+\mu} \psi_{-\mu} 
  - \chi_{+i} \chi_{-i} -\frac{1}{4} \eta_{+} \eta_{-} 
\right\},
\label{eq:exact_cont_action}
\end{eqnarray}
where
\begin{eqnarray}
F_{0i} = D_0 A_i,\qquad  F_{ij} = i[A_i,A_j],   \qquad  (i,j=1,2,3).
\label{field_tensor}
\end{eqnarray} 
These $F_{\mu\nu}$ are originally the field tensors in four dimensions and  become the covariant derivatives and the commutators after the dimensional reduction.

From the nilpotency  (\ref{eq:q2}), the action (\ref{eq:exact_cont_action}) is trivially invariant under the $Q_{\pm}$-transformations, without a usage of the Leibniz rule.
The other fourteen supersymmetries and the global $SO(9)$ symmetry are also exact symmetries of (\ref{eq:exact_cont_action})
 because we only performed the change of variables from the original variables to the twisted ones.

\subsection{Lattice theory}
\label{lattice_theory}
%
%
% Introduction to lattice 
We construct the theory on a finite lattice of size $L$ with periodic boundary condition.
The sites of the lattice are labeled by the integers $t=1,\cdots,L$ and 
the lattice spacing $a$ is set to unity without loss of generality.
The adjoint scalars and fermions live on the sites, while a lattice gauge field $U(t) \in SU(N)$ lives on 
the links.% and will be referred to as the link field in the following. 

The lattice gauge transformations are given by
\begin{eqnarray}
\delta_\omega U(t) =- i\nabla_0 \, \omega(t) \, U(t),
\qquad \delta_\omega \varphi(t) = -i [\varphi(t),\omega(t)],  
\label{def:lattice_gauge_tr}
\end{eqnarray}
where $\varphi$ represents all of the scalars and fermions,
and $\omega$ is an infinitesimal gauge parameter defined on the sites.
Under the gauge transformations, 
the covariant forward difference operator 
\begin{eqnarray}
\nabla_0 \varphi(t) = U(t) \varphi(t+1) U^{-1}(t) - \varphi(t)
\label{cov_dif_op}
\end{eqnarray}
transforms as $\varphi(t)$ itself.
%for adjoint fields $\varphi$.
%
%

By replacing the integral and the covariant derivative in (\ref{eq:exact_cont_action}) and (\ref{field_tensor})
with the summation over the sites and the covariant forward difference operator (\ref{cov_dif_op}), respectively,  
we can introduce the following lattice action,
\footnote{ 
This is a one dimensional version of the action given in \cite{Sugino:2004uv},
 which is therefore referred to as the Sugino lattice action in this paper.
}
\begin{eqnarray}
S
& =&
  Q_{+} Q_{-} \frac{N}{2 \lambda_0} \sum_{t=0}^{L-1} \; {\rm tr} 
  \left\{
            -2i B_{i}  \left( F_{i3} + \frac{1}{2} \epsilon_{ijk}F_{jk} \right)
            -\frac{1}{3} \epsilon_{ijk} B_{i}[B_{j},B_{k}]
  \right. 
  \nonumber
\\
&& \hspace{3.8cm}
\left.
  - \psi_{+\mu} \psi_{-\mu} 
  - \chi_{+i} \chi_{-i} -\frac{1}{4} \eta_{+} \eta_{-} 
\right\},
\label{eq:lat_action}
\end{eqnarray}
where
\begin{eqnarray}
&& F_{0i} = \nabla_0 A_i,\qquad  F_{ij} = i[A_i,A_j],  \qquad  (i,j=1,2,3).
\end{eqnarray}
The  't Hooft coupling $\lambda$ has mass dimension three, and  therefore  $\lambda_0=\lambda a^3$ is
the dimensionless 't Hooft coupling constant.
The continuum limit of this theory is obtained by letting $\lambda_0$ to zero with fixed $\lambda$.

The lattice $Q_\pm$-transformations
\begin{eqnarray}
\begin{array}{ll}
   Q_{\pm} U = i \psi_{\pm 0} U,
&  \hspace{1cm}
\nonumber \\ 
   Q_{\pm} A_\mu = \psi_{\pm \mu},
&  \hspace{1cm}
   Q_{\pm} \psi_{\pm \mu} = -i \nabla_\mu \phi_{\pm} + i\delta_{\mu 0} \psi_{\pm 0}\psi_{\pm 0},
\nonumber \\ 
   Q_{\pm} B_i   = \chi_{\pm i}, 
&  \hspace{1cm} 
   Q_{\pm} \chi_{\pm i} = [B_i, \phi_{\pm}], 
\nonumber \\
   Q_{\pm} C     = \eta_{\pm},
&  \hspace{1cm}  
   Q_{\pm} \eta_{\pm} = [C, \phi_{\pm}],
\nonumber \\
   Q_{\pm} \phi_{\mp} = \eta_{\mp},
&  \hspace{1cm}  
   Q_{\pm} \eta_{\mp} = [\phi_\mp, \phi_{\pm}], 
\nonumber \\
   Q_{\pm} \phi_{\pm} = 0,
&
\nonumber \\
&  \hspace{1cm}  
   Q_{\pm} \chi_{\mp i} = \frac{1}{2}[C,B_i] \pm H_i,
\nonumber \\
&  \hspace{1cm}  
    Q_{\pm} \psi_{\mp \mu} = \frac{i}{2} \nabla_\mu C 
                              \pm \tilde H_\mu 
                            +   \frac{i}{2}\delta_{\mu 0} \{\psi_{+0},\psi_{-0}\}, 
\nonumber \vspace{-1mm}
\end{array}
\end{eqnarray}
\begin{eqnarray}
&&   Q_{\pm} H_i = \pm \left(              [ \chi_{\mp i}, \phi_\pm ]
                     + \frac{1}{2} [ \chi_{\pm i}, C        ]
                     + \frac{1}{2} [ B_i,      \eta_\pm ] \right), \hspace{10mm}
\nonumber \\
&&   Q_{\pm} \tilde H_\mu =  \pm \left(              [ \psi_{\mp \mu}, \phi_\pm ]
                               + \frac{1}{2} [ \psi_{\pm \mu}, C        ]
                               - \frac{i}{2} \nabla_\mu \eta_\pm 
                               \right)
\label{lattice_Q}
                               \\
&& \hspace{2cm} \pm \delta_{\mu 0} 
\left( \frac{1}{4}
   \left[ \psi_{\pm 0}, \, \nabla_0 C \pm 2i  H + \{\psi_{+0},\psi_{-0}\} \right]
    +\frac{1}{2} [\psi_{\mp 0},\nabla_0 \phi_{\pm}] 
\right)
\nonumber 
\end{eqnarray}
are the same with the continuum ones with the exception of the fields with $\mu=0$.
The terms proportional to $\delta_{\mu0}$ are higher order corrections.
These  $Q_\pm$  satisfy 
\begin{eqnarray}
Q_{\pm}^{2}= i \delta_{\phi_\pm}, \quad \{ Q_{+},Q_{-} \} = -i \delta_{C},
\label{eq:lat_2q}
\end{eqnarray}
even on the lattice \cite{Sugino:2004uv}. There is, however, nothing surprising about this if we recall the BRST transformation.
In the BRST transformation, once the transformation law of a low dimensional field is given, 
then that of the higher dimensional field is uniquely determined from the nilpotency. 
Similarly, even on the lattice, once the $Q_\pm$-transformation laws of the low dimensional fields ($Q_\pm U$, $Q_\pm B_i$, $etc$.) are given,
then the others ($Q_\pm \psi$, $Q_\pm \chi_\pm$, $etc$.) are uniquely determined from (\ref{eq:lat_2q}).

The lattice action  (\ref{eq:lat_action})  possesses  the $Q_{\pm}$-invariance and the gauge invariance.
On the other hand, the other fourteen supersymmetries and the $SO(9)$ global symmetry are broken at a finite lattice spacing.
Instead of the $SO(9)$ symmetry,  the lattice theory has,  at least, a $SU(2)$ global symmetry corresponding to  the interchange of $Q_+$ and $Q_-$.
In the naive continuum limit,  the lattice $Q_\pm$-transformations (\ref{lattice_Q}) reproduce the continuum ones (\ref{cont_Q}) and the lattice action reproduces the correct continuum action (\ref{def:1dSYM_action}) via (\ref{eq:exact_cont_action}). 
%
%
%
%

%Now, let us write the lattice action in terms of original variables.
After performing $Q_{\pm}$ in (\ref{eq:lat_action}), 
we find that the lattice action has a
four-fermi interaction,
\begin{eqnarray}
  S_{4f}=\frac{N}{2\lambda_0}\sum_t   \mathrm{tr} \left( -\frac{1}{4} \left\{\psi_{+0}, \psi_{-0} \right\}^2 \right),
  \label{four_fermi}
\end{eqnarray}
which is  of the order of the cut-off. 
The four-fermi term is not suited for the numerical simulations.
In order to express the four fermion interaction as a fermion bilinear form,
as in the NJL model,
we introduce an auxiliary field $\sigma(t)$ and replace (\ref{four_fermi}) with
\begin{eqnarray}
S_{4f} =
\frac{N}{2\lambda_0}\sum_t \mathrm{tr} \big\{ \sigma^2 +  \psi_{+0} [\sigma, \psi_{-0} ] \frac{}{}  \big\}.
\end{eqnarray}
In fact, this reproduces (\ref{four_fermi}) by integrating $\sigma$.

\section{Simulation details}
\label{simulation_details}

We used the standard rational Hybrid Monte Carlo method for the numerical simulations \cite{Clark:2004cp}.
As explained in section \ref{lattice_theory}, the auxiliary field $\sigma$ was introduced
to express the fermi action as a fermion bilinear from.
Therefore, the HMC generated the configurations of the eleven bosonic variables, $U,X_1,\cdots,X_{9}$ and  $\sigma$.
Moreover, as explained below, we prepared the fermion pfaffian by using the pseudo fermion method with a rational approximation. 
For the phase of the pfaffian, 
we used the phase reweighting method  for the SUSY WTI  in section \ref{sec:sim_res} and quenched it for the internal energy of the black hole in section \ref{internal_energy}.
\footnote{ 
This is purely due to some technical reasons. 
The numerical calculation of the pfaffian phase is demanding and  costs $O(K^3)$
for $K \times K$ matrices. 
We have to use sufficiently large $N$ for testing the gauge gravity duality in the large $N$ limit 
and avoiding the instability related to the flat directions. 
If we use a lattice of size $L=16$ and $N=16$, then  the size of the Dirac operator becomes
$K=(N^2-1) \times 16\times L \simeq 65,000$, which is too large for the calculation of the phase.
Thus, we did not compute the pfaffian phase when we measured the internal energy of the black hole.
On the other hand, $N=2,3,4$ are sufficient to confirm the restoration of SUSY from the SUSY WTI. 
For such small $N$, we computed the phase every configuration.
}

%For supersymmetric theories, the quench approach is absurd 
%because bosons and fermions have to be treated equality under supersymmetry.
The dynamical effects of the fermions can be included in the simulations through the pseudo fermion method.
In the present model, 
the integration of the fermions yields the pfaffian ${\rm pf}(D)$ which is complex values in general.  
We treated the absolute value and the complex phase of the pfaffian, individually. 
Since $|{\rm pf}(D)|={\rm det}(D^\dag D)^{1/4}$, 
the absolute value of the pfaffian can be given by
\begin{eqnarray}
\displaystyle
|{\rm pf}(D)| = \int D\phi^\dag D \phi\, {\rm exp}\left\{
- \phi^\dag (D^\dag D)^{-\frac{1}{4}}  \phi
\right\},
\end{eqnarray}
where  $\phi$ is a  complex pseudo fermion,
and the 4th-root of $D^\dag D$ can be approximately prepared by a rational expansion,   
\begin{eqnarray}
(D^\dag D)^{-\frac{1}{4}}   = \alpha_0 + \sum_{i=1}^M \frac{\alpha_i}{D^\dag D + \beta_i}, 
\label{eq:rational_approx}
\end{eqnarray}
where the degree $M$ and the coefficients $\alpha_i,\beta_i$
depend on the range of the eigenvalues of $D^\dag D$ and the accuracy of the approximation.
We examined the range of the eigenvalues at the thermalization step in each simulation parameter
and chose $M,\alpha_i,\beta_i$ of the approximation within
the relative errors of ${\cal O}(10^{-14})$ \cite{Clark:2005int}.
Then, we fixed them during the configuration generation.
We justified our initial guess for $M,\alpha_i,\beta_i$
by computing the maximum and minimum eigenvalues in each trajectory. 
In addition, we computed the inversions of $D^\dag D$  with the shifts $\beta_i$ in (\ref{eq:rational_approx}) 
by using the multi-mass CG solver \cite{Frommer:1995ik}. 
The effect of the phase of the pfaffian was included in the results of the SUSY WTI using the phase reweighting method.
Meanwhile, we merely ignored it when we measured the internal energy of the black hole.

As already reported in \cite{Anagnostopoulos:2007fw}, \cite{Catterall:2008yz}, \cite{Kanamori:2008bk},
the HMC achieved the thermalization at high temperature, 
while at low temperature it was unstable and we observed that the magnitudes of the scalar fields  have been monotonically increasing
against the trajectories. 
Once such phenomena occurred, the thermalization was not achieved. 
\footnote{ The  threshold temperature, where we could observe such phenomena, depended on $N$,  and the larger $N$ were more stable.  
This instability comes from the flat directions of the boson action.} 
Therefore, in the parameter regions with the instability,
we added a mass term for the scalar fields,  
\begin{eqnarray}
S_{\mathrm{mass}} &=& \mu^{2}_{0}\, \frac{N}{2\lambda_0}\,  \sum_{t=1}^{L} \; \sum_{i=1}^{9} 
 \mathrm{tr} \left( X^{2}_{i} (t) \right), \label{eq:mass}
\end{eqnarray}
to the action, where $\mu_0(=\mu a)$ is the dimensionless mass. 
This term was used to examine the SUSY restoration as presented in the next section. 
Meanwhile, in section \ref{internal_energy}, we measured the internal energy of the black hole without the mass term 
by taking sufficiently large $N$ to avoid the instability.

The trajectory length was $\tau=0.5$,  and the number of step of the molecular dynamics was chosen such that the acceptance rate was between 80\% and 90\%.
After discarding the first ${\cal O}(10^3)$ trajectories for the thermalization, we stored 1 configuration every 10 trajectories.\\

%%%%%%%%%%%%%%%%%%%%%%%%%%%%%%%%%%%%%%%%%%%%%%%%%%%%%%%%%%%%%%%%%%%%%%%%%%%%%%%%%%%%%%%%%%%%%%%%%%%%
%                                                                                                                                                   %
%                                                                                                                                                   %
%              Restoration of supersymmetry                                                                                              %
%                                                                                                                                                   %
%                                                                                                                                                   %
%%%%%%%%%%%%%%%%%%%%%%%%%%%%%%%%%%%%%%%%%%%%%%%%%%%%%%%%%%%%%%%%%%%%%%%%%%%%%%%%%%%%%%%%%%%%%%%%%%%%

\section{Restoration of supersymmetry} \label{sec:sim_res}

In this section, we see that the effect of the lattice spacing vanishes and supersymmetry is restored in the continuum limit by numerically examining the supersymmetric Ward-Takahashi identity with the SUSY breaking term (partially conserved supercurrent).

The present lattice action has two exact supercharges, but the remaining fourteen supercharges 
are broken by the lattice cut-off. 
In the classical continuum limit, the broken supersymmetries are trivially restored, while
in the quantum theory the restoration does not occur in general because
 some SUSY breaking relevant operators can be generated by the UV-divergences.
Fortunately, in one dimensional gauge theories, the perturbative power-counting tells us that no SUSY breaking relevant operators exist, in other words, the theories are UV-finite.
Hence, the supersymmetries are restored in the quantum continuum limit. 
However, it is not clear whether this argument surely holds beyond the perturbation theory and at finite temperature. 
Therefore, we need to show that the restoration does occur.

\subsection{Supersymmetric Ward-Takahashi identity}

In the following, we derive the supersymmetric Ward-Takahashi identity in the continuum theory at finite temperature using the path integral formulation.

We define the partition function and the expectation value of an operator $O$ as follows:
\begin{eqnarray}
Z &=& \int D\varphi \  \  {\rm e}^{ -S'_{cont.}},  
\label{eq:partition}\\
\left\langle \mathcal{O}\right\rangle &=& \frac{1}{Z}\int D\varphi \  \mathcal{O}(\varphi) \  {\rm e}^{ -S'_{cont.}}, \label{eq:expection}
\end{eqnarray}
where $\varphi$ represents the whole field, and the continuum action $S'_{cont.}$ is given by
\begin{eqnarray}
S'_{cont.} = S_{cont.} + \frac{N \mu^{2}}{2\lambda}\int dt \,  \sum_{i=1}^{9} {\rm tr}\left\{X_{i}^2(t) \right\}.
\end{eqnarray}
We added the mass term to the action.
Hereafter, we take a fermionic one point operator ${\cal O}_\beta(s)$ as ${\cal O}$.

For the integration variables, we make a change of variables, which is given by the localized supersymmetry transformations, 
(\ref{def:1d_susy}) with $\xi \rightarrow \xi(t)$.
Note that the local parameter $\xi(t)$ satisfies the anti-periodic boundary condition. 
Under the change of variables, $S_{cont.}$ changes to the total derivative of the supercurrent,
\begin{eqnarray}
J_{\alpha } &=& \frac{N}{\lambda} 
 \left[
   (\gamma_i)_{\alpha\beta} \ \mathrm{tr} \left(\psi_\beta D_0 X_i \right)
   -i (\Sigma_{ij})_{\alpha\beta} \  
     \mathrm{tr} ( \psi_{\beta} [X_i,X_j] ) 
 \right], 
 \label{cont_current}
\end{eqnarray} 
and the mass term yields a supersymmetry breaking term,
\begin{eqnarray}
Y_\alpha &=& \frac{N}{\lambda} 
(\gamma_i)_{\alpha\beta} \
\mathrm{tr} \left\{ X_i \psi_\beta \right\}. \label{eq:Y}
\end{eqnarray}
The integral measure is invariant under the change. Thus, we find
\begin{eqnarray}
\int dt \ \xi_{\alpha} (t) \frac{d}{dt}  \left\langle J_{\alpha}(t) \mathcal{O}_{\beta} (s) \right\rangle &=& 
\mu^{2} \int dt \  \xi_{\alpha}(t) \left\langle Y_{\alpha} (t) \mathcal{O}_{\beta} (s) \right\rangle \nonumber \\
& & -\int dt \,  \delta(t-s) \,  \xi_{\alpha}(t) \left\langle Q_{\alpha} \mathcal{O}_{\beta}(s)   \right\rangle, \label{eq:wtid}
\end{eqnarray}
where the supercharges $Q_\alpha$ in the last term give the infinitesimal supersymmetry transformations.
Since $\xi$ in (\ref{eq:wtid}) is arbitrary, one can obtain the SUSY Ward-Takahashi identity,
\begin{eqnarray}
\frac{d}{dt}\left\langle J_{\alpha} (t) \mathcal{O}_\beta(s) \right\rangle 
&=& \mu^{2} \left\langle  Y_{\alpha} (t)  \mathcal{O}_\beta(s) \right\rangle 
- \delta(t-s) \left\langle  Q_{\alpha} \mathcal{O}_\beta(s) \right\rangle
\label{eq:pcsc}
\end{eqnarray}
This derivation of the SUSY WTI is independent to the boundary condition, hence we find that (\ref{eq:pcsc}) holds for both periodic and anti-periodic boundary conditions.
From a lattice version of the supersymmetric Ward-Takahashi identity, 
one can investigate the supersymmetry breaking effect of the lattice spacing, as distinguished from those of the temperature and the mass \cite{Kanamori:2008bk}.

\subsection{Numerical results of SUSY WTI}

In this section, we present the numerical results of the SUSY WTI,  including
the effect of the complex phase of the pfaffian, for $SU(2)$ and  at the temperature $T/\lambda^{1/3}=1$.
%The results for $SU(3)$ and $SU(4)$ are the same as those of $SU(2)$ and are summarized in the appendix. 
The parameters set used in the simulations is presented in table \ref{tab:parameters}.
\begin{table}
\begin{center}
\begin{tabular}{|ccccccccc|}
\hline
\multicolumn{9}{|c|}{SU(2)     $\quad T/\lambda^{1/3}=1$} \\ 
\noalign{\global\arrayrulewidth 1.0pt}
\hline \hline
\noalign{\global\arrayrulewidth.4pt}
&$L$ && $a$ && $\mu^{2}$ && traj. &\\
\hline
& 8 & & 0.125 & & 0.01 & & 15,000 & \\
& 8 & & 0.125 & & 0.02 & & 15,000 & \\
& 8 & & 0.125 & & 0.05 & & 15,000 & \\
& 12 & & 0.0833 & & 0.01 & & 15,000 & \\
& 12 & & 0.0833 & & 0.02 & & 15,000 & \\
& 12 & & 0.0833 & & 0.05 & & 15,000 & \\
& 16 & & 0.0625 & & 0.01 & & 15,000 & \\
& 16 & & 0.0625 & & 0.02 & & 15,000 & \\
& 16 & & 0.0625 & & 0.05 & & 15,000 & \\
\hline
\end{tabular}
%}
\vspace{3mm}
\caption{Simulation parameters and the HMC trajectories. 
The lattice spacing $a=(TL)^{-1}$ and the mass $\mu^{2}$ are presented in the  physical unit ($\lambda=1$). 
\vspace{3mm}
}
\label{tab:parameters}
\label{tab:traj-tab}
\end{center}
\end{table}

At this temperature, we observed the instability related to the flat directions.
Therefore,  as explained in section \ref{simulation_details}, we added the mass term to the lattice action 
and measured the SUSY WTI with the breaking term $Y$.
For the simplicity of the explanation, we give the results in the physical unit ($\lambda=1$).

As a lattice counterpart  of the SUSY WTI, we chose
$Y$ (\ref{eq:Y}) as the source operator ${\cal O}$ and calculated the following sixteen ratios,
\footnote{ In fact, we measured  the ratios,
 \begin{eqnarray}
\frac{ \left\langle d J_{\alpha}(t) Y_\beta(0) \right\rangle }{ \left\langle Y_\alpha(t)Y_\beta(0) \right\rangle }, \qquad {\rm for}\ \ \alpha, \beta=1.\cdots,16, \label{eq:full_ratio}
\end{eqnarray}
In the present notation of the gamma matrices, the ratios (\ref{eq:full_ratio}) with $\alpha \neq \beta$ were zero within the statistical errors. }
\begin{eqnarray}
\frac{ \left\langle d J_{\alpha}(t) Y_\alpha(0) \right\rangle }{ \left\langle Y_\alpha(t)Y_\alpha(0) \right\rangle }, \qquad {\rm for}\ \ \alpha=1,\cdots,16, \label{eq:ratio}
\end{eqnarray}
where $d$ is the symmetric difference operator, $ dJ_\alpha(t)=[J_\alpha(t+1)-J_\alpha(t-1)]/2$, and 
the lattice supercurrents $J_\alpha$ are prepared by replacing the covariant derivative in (\ref{cont_current}) with the covariant forward difference operator. 
These sixteen ratios correspond to the sixteen supercharges.
Note that there are two exact supercurrents corresponding to $Q_{\pm}$. Since their concrete forms are complicated,
we used this simple definition for all of the supercharges.
The calculation of the inversion of the Dirac operator is needed to estimate the ratios.
We used LAPACK \cite{lapack}  to compute the inversion. 
Averaging the ratios over the lattice sites, we reduced the statistical errors.

%Fig.\ref{} shows the sharps of the correlators $JY, YY$ and $dJY$ for $\mu^{2}=0.01$.
%The statistical errors are estimated by the standard Jackknife method, and are about *****\% respectively.

%%%%%%%%%%% SU(2) %%%%%%%%%%
\begin{figure}[htbp]
 \begin{center}
  \includegraphics[width=115mm]{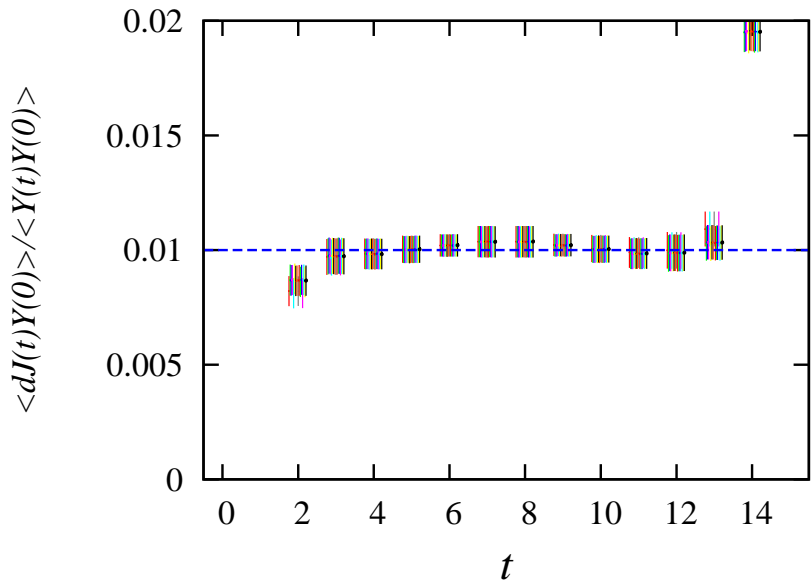}
 \end{center}
  \caption{Ratios related to the SUSY WTI at $T=1$ for $L=16$ and $\mu^2=0.01$ (dashed blue line) in the $SU(2)$ theory. 
  The horizontal axis is time on which the supercurrents are defined, while the source operator is put on $t=0$. 
  \vspace{10mm}
    }
 \label{fig:N2L16T1M1}
\end{figure} 

Figure \ref{fig:N2L16T1M1} shows the ratios against time on which the supercurrents are defined for $L=16$ and $\mu^{2}=0.01$.
The corresponding lattice spacing is $a=0.0625$.
There are sixteen points corresponding to the sixteen supercharges on the same time-slice.
Clear plateaus are seen for $3 \le t \le 13$.  
All of the sixteen plateau values are $\mu^2=0.01$ within the errors. In other words, the SUSY WTI holds near the continuum limit.

On the other hand, near $t=0$ (and $L$), the ratios deviate from the plateaus because of the contact term in (\ref{eq:pcsc}). 
The contact terms are smeared on the lattice, therefore the ratios 
near $t=0$ (even for $t \neq 0$) can show large deviations. 
We also found that the ratios with different spinor indices were well degenerate.  
In the continuum theory, the sixteen supercurrents are interchanged with each other under the $SO(9)$ transformations.
The $SO(9)$ internal symmetry is broken to $SU(2)$ in the present lattice model.
The degeneracies indicate that the broken $SO(9)$ symmetry is mostly restored at this lattice spacing. 

We estimated the plateau values using the constant $\chi^{2}$-fit, after which we extrapolated the fitted values to the continuum limit.
Figure \ref{fig:N2data_10-M} shows the extrapolated values for three different masses, $\mu^{2}=0.01,0.02,0.05$.
For all of the three masses, the results were about $\mu^2$ within the statistical errors. 
Thus, we found that the SUSY WTI holds in the continuum limit and the lattice theory surely reproduces the correct continuum theory.

Finally, we examine the massless limit.
For the extrapolation to the massless limit, we used the following linear function,
\begin{eqnarray}
f(\mu^{2})=c \mu^{2} + d,
\end{eqnarray}
where $c,d$ are the fitting parameters. In table \ref{fig:N2data_10-M}, $c$ and $d$ are presented for each $\alpha$.  
For all of the spinor indices, $c=1$ and $d=0$ within the errors.
These results provide the possibility that the SUSY theory can be defined by using the extrapolation to the massless limit
even for the parameter region where the instability is observed. \\

%
%\begin{figure}[t]
%\begin{minipage}{0.5\hsize}
%\begin{center}
%\hspace{-0.6cm}
%\includegraphics[width=7cm,keepaspectratio,clip]{./figs/su2jx_10-001.eps}
%\end{center}
%%
%\end{minipage}
%\begin{minipage}{0.5\hsize}
%\begin{center}
%\includegraphics[width=7cm,keepaspectratio,clip]{./figs/su2djx_10-001.eps}
%\end{center}
%\end{minipage}
%%
%\begin{minipage}{0.5\hsize}
%\begin{center}
%\includegraphics[width=7cm,keepaspectratio,clip]{./figs/su2xx_10-001.eps}
%\end{center}
%\end{minipage}
%%
%\begin{minipage}{0.5\hsize}
%\begin{center}
%\end{center}
%\end{minipage}
%%
%\caption{Behavior of the corelators $JY$(Left), $\partial^{s}_{0}JY$(Center), and $YY$(Right)
% with $\alpha=1,5,9,13$ in SU(2).
% The squared masses and lattice site is  chosen as $\mu^{2}=0.01$, $L=16$.}
%\label{fig:jy}
%\end{figure}

%%%%% SU(2) %%%%%
\begin{figure}[ht]    
 \begin{center}
  \includegraphics[width=100mm]{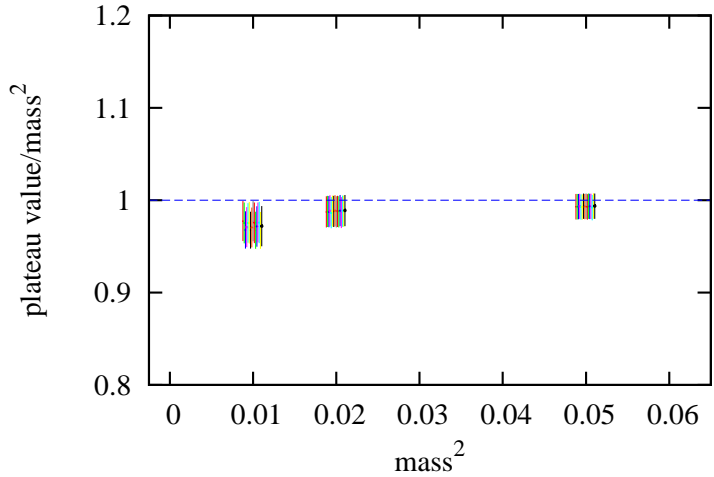}
 \end{center}
 \caption{Mass dependence of the plateau values at $T=1$ in $SU(2)$. The $x$-axis is $\mu^2$ and 
 the $y$-axis represents the plateau values, extrapolated to the continuum limit, over the mass squared. }
 \label{fig:N2data_10-M}
 \vspace{15mm} 
 %\label{fig:one}
\end{figure}

\begin{table}[htb]
\begin{center}
  \begin{tabular}{cccccccccc}
\hline \hline
$\alpha$  & &  $c$ & & & $d$ & & & $\chi^{2}$/dof & \\ \hline \hline
 1  & & 0.997(17)(12) & & & -0.00019(32)(17) & & & 1.7e-04 &  \\ 
 2  & & 0.998(17)(13) & & & -0.00022(32)(18) & & & 4.6e-03 &  \\ 
 3  & & 1.001(17)(15) & & & -0.00031(30)(18) & & & 3.9e-02 &  \\ 
 4  & & 1.000(17)(12) & & & -0.00027(32)(21) & & & 3.0e-02 &  \\ 
 5  & & 0.997(17)(13) & & & -0.00021(32)(18) & & & 3.8e-04 &  \\ 
 6  & & 0.997(17)(12) & & & -0.00021(33)(19) & & & 9.7e-03 &  \\ 
 7  & & 1.001(17)(14) & & & -0.00032(30)(19) & & & 3.9e-02 &  \\ 
 8  & & 1.000(17)(12) & & & -0.00028(32)(20) & & & 3.7e-02 &  \\ 
 9  & & 0.997(17)(12) & & & -0.00020(32)(17) & & & 1.1e-05 &  \\ 
 10  & & 0.997(17)(14) & & & -0.00021(32)(18) & & & 6.2e-03 &  \\ 
 11  & & 1.001(17)(14) & & & -0.00032(30)(18) & & & 3.4e-02 &  \\ 
 12  & & 1.000(17)(12) & & & -0.00027(32)(20) & & & 2.9e-02 &  \\ 
 13  & & 0.998(17)(13) & & & -0.00022(32)(17) & & & 9.1e-05 &  \\ 
 14  & & 0.997(17)(13) & & & -0.00020(33)(19) & & & 1.1e-02  & \\ 
 15  & & 1.001(17)(14) & & & -0.00033(30)(18) & & & 4.4e-02  & \\ 
 16  & & 1.000(17)(12) & & & -0.00027(32)(20) & & & 2.7e-02  & \\ 
\hline \hline 
  \end{tabular}
  \vspace{5mm}
  \caption{Fit results for the plateau values using the linear function, $f(\mu^{2})=c \mu^{2} + d$.}
 \label{fig:N2data_10-M}
\end{center}
\end{table}

%%%%%%%%%%%%%%%%%%%%%%%%%%%%%%%%%%%%%%%%%%%%%%%%%%%%%%%%%%%%%%%%%%%%%%%%%%%%%%%%%%%%%%%%%%%%%%%%%%%%
%                                                                                                                                                   %
%                                                                                                                                                   %
%             Internal energy of the black hole                                                                                            %
%                                                                                                                                                   %
%                                                                                                                                                   %
%%%%%%%%%%%%%%%%%%%%%%%%%%%%%%%%%%%%%%%%%%%%%%%%%%%%%%%%%%%%%%%%%%%%%%%%%%%%%%%%%%%%%%%%%%%%%%%%%%%%

\section{Internal energy of the black hole}
\label{internal_energy}
The internal energy of the black hole is a fundamental quantity to study the duality conjecture for the present theory.
In this section, 
we see the numerical result of the internal energy, where the effect of the phase of the fermion pfaffian is quenched.
By comparing them with the theoretical prediction of the gravity side, we examine the validity of the conjecture.
The numerical results of the internal energy are also found in
\cite{Anagnostopoulos:2007fw}, \cite{Hanada:2008ez}, \cite{Hanada:2013rga},
\cite{Catterall:2007fp}, \cite{Catterall:2008yz}, \cite{Catterall:2009xn}.

The gravity side predicts that the internal energy of the non-extremal black hole, which is the gravity dual of the target gauge theory,
  behaves
\begin{eqnarray}
\frac{1}{N^2}
\left( \frac{E}{\lambda^{1/3}} \right)=c_1 
\left( \frac{T}{\lambda^{1/3}}
\right)^{2.8} 
+c_2 \left( \frac{T}{\lambda^{1/3}} 
\right)^{4.6}+ \cdots,
\label{energy_nlo}
\end{eqnarray}
in the large $N$ limit. The leading order coefficient $c_1$ can be calculated by hand \cite{Klebanov:1996un}:
\begin{eqnarray}
c_1=
\frac{9}{14} \left(4^{13} 15^2 \left(\frac{\pi}{7}\right)^{14}\right)^{\frac{1}{5}} 
=
7.407....
\end{eqnarray}
In contrast,  the next-to-leading order coefficient $c_2$ is unknown. 
%The  $\alpha'$-corrections of string theory give $c_2$  in principle. 

In the gauge theory side, one can estimate the internal energy of the gauge theory from the expectation value of the boson action
\cite{Catterall:2007fp}. 
Indeed,  for the definition of the internal energy $ \langle E\rangle \equiv -\frac{\partial}{\partial \beta} {\rm log}Z$ with $\beta=1/T$, we can perform the $\beta$ derivative of ${\rm log} Z$ analytically and obtain
\begin{eqnarray}
\langle E\rangle =-\frac{3}{\beta} \langle S
\rangle,
\label{internal_energy_in_gauge_side1}
\end{eqnarray}
where $S$ is given by (\ref{eq:exact_cont_action}).
The $Q$-exactness of the total action implies that the internal energy (\ref{internal_energy_in_gauge_side1}) vanishes in the zero temperature limit where the effect of the SUSY-breaking boundary conditions fade away.

On the lattice, we define the internal energy using (\ref{internal_energy_in_gauge_side1}) 
with the lattice action (\ref{eq:lat_action}). 
Since the integrations of the fermions and the seven auxiliary fields can be analytically performed,  it turns out to be
\begin{eqnarray}
\langle E\rangle = -\frac{3}{\beta} \left\{  \langle S_B
\rangle 
-\frac{9+k}{2} N_t(N^2-1)
\right\},
\label{internal_energy_in_gauge_side2}
\end{eqnarray}
where $k$ is the number of unintegrated auxiliary fields, and if $k \neq 0$ the boson action $S_B$ includes the action of the $k$ auxiliary fields.
As explained in section \ref{simulation_details}, the present lattice action has the auxiliary field $\sigma$ 
which remains unintegrated so that $k=1$. 
Using (\ref{internal_energy_in_gauge_side2}) with the lattice boson action $S_B$ extracted from (\ref{eq:lat_action})
%(\ref{lattice_boson_action})
 and $k=1$,  
one can measure the internal energy from the lattice simulations.

\begin{figure}[htbp]
 \begin{center}
  \includegraphics[width=115mm]{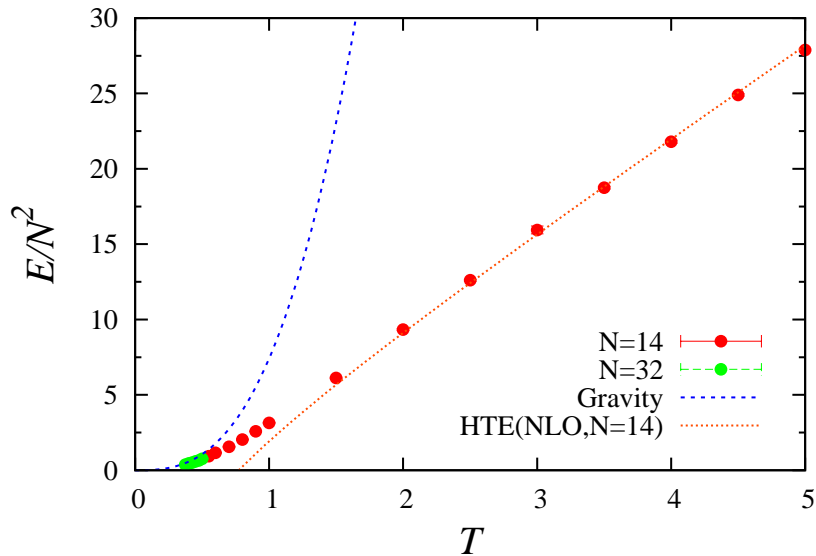}
   \caption{Internal energy of the black hole against temperature. The simulation results (red for $N=14$, green for $N=32$) coincide with the result of the high temperature expansion (dashed orange curve) at high temperature and  approach the theoretical prediction (dashed blue curve) as the temperature decreases.  
   \vspace{10mm}
   }
   \label{fig:internal_energy}
 \end{center}
\end{figure}

\begin{figure}[htbp]
 \begin{center}
  \includegraphics[width=115mm]{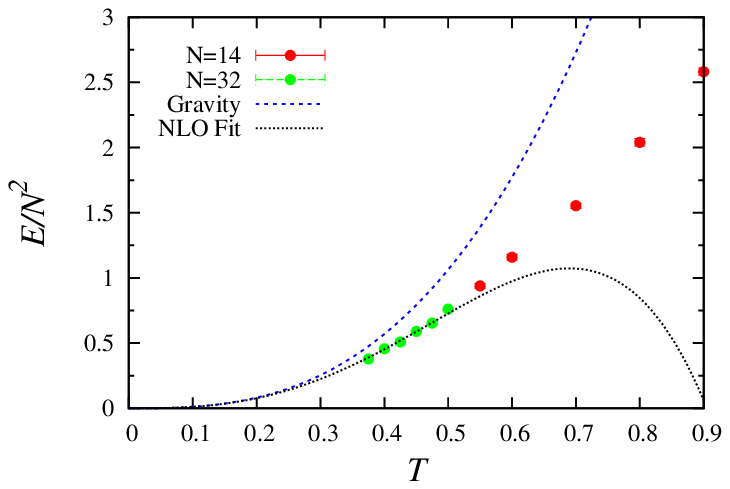}
   \caption{Low temperature region of the internal energy of the black hole.  The dashed blue curve is again the theoretical prediction of the gravity side up to the leading order. The dashed curve denotes the fit result which is obtained by fitting $5$ points within $T=0.375-0.475$.  }
   \label{fig:internal_energy_lowT}
 \end{center}
\end{figure}

Figure \ref{fig:internal_energy} shows the lattice results of the internal energy,  where the red and green points denote the results for $N=14$ and ones for $N=32$, respectively. The lattice size is $L=16$ for $T<1$ and $L=8$ for $1 \le T \le 5$ with the corresponding lattice spacing $aT_{\rm eff}=1/L$.
The $x$-axis denotes dimensionless temperature $T_{\rm eff}=T/\lambda^{1/3}$. 
The dashed blue curve represents the theoretical prediction of the gravity side at the leading order of the expansion,
 (\ref{energy_nlo}) with $c_2=0$. The slightly curved orange line is the result of the high temperature expansion \cite{Kawahara:2007ib}, \cite{kadoh_hte}.
As can be seen in the figure, the lattice data coincide with the result of the high temperature expansion at high temperature, while as  the temperature decreases, %apart from the line of the HTE, 
they smoothly approach the theoretical prediction of the gravity side.

In figure \ref{fig:internal_energy_lowT}, we focus on low temperature of figure \ref{fig:internal_energy}. 
The data surely approach the prediction (dashed blue curve)  and are likely to coincide with it as the temperature decreases further.
But, unfortunately, the temperatures we used in the simulations were not low enough to explain the leading behavior of the gravity side.
To obtain quantitative results for the leading-order term, simulations at further low temperatures are required.
%to and no quantitative results for the leading-order term has been obtained so far.

Instead, one can study the contribution of the next-to-leading order term by fitting the lattice results using the following formula,
\begin{eqnarray}
f(x)= 7.41 x^{2.8} + C x^p,
\label{fit_formulta__nlo}
\end{eqnarray}
where $C$ and $p$ are the fitting parameters. 
From (\ref{energy_nlo}), 
if the duality conjecture is really true, the obtained $p$ should be $4.6$ within the statistical errors. 
We performed the fit using the 5 points within $0.375 \le T \le 0.475$ and obtained
\begin{eqnarray}
C=9.0 \pm 2.6,  \qquad p=4.74 \pm 0.35.
\label{fit_result_energy_nlo}
\end{eqnarray}
The obtained p is consistent with the theoretical prediction of the gravity side within about seven percent statistical error. 
This is the first lattice result of the NLO term, which quantitatively shows the validity of the duality conjecture in this system.

In \cite{Hanada:2008ez}, the NLO term was estimated from the numerical simulation 
based on the momentum sharp cut-off method using the same fit formula (\ref{fit_formulta__nlo}).
The fitted values $C=5.55(7)$, $p=4.58(3)$ are consistent with our results (\ref{fit_result_energy_nlo}) within two sigma. 
However, those values were obtained from their data in a little high temperature region $0.5 \le T \le 0.7$. 
We also tried to fit the lattice results in the same temperature region,
but could not obtain a reasonable result within $\chi^2/dof \lesssim 1$.
This observation raised the possibility that the temperature region $0.5 \le T \le 0.7$ was a little high to estimate the next-leading order term.
The reason of the discrepancy between the two results have not  been clear in detail so far. 
Further simulations are now in progress and will give us the final answer.

%%%%%%%%%%%%%%%%%%%%%%%%%%%%%%%%%%%%%%%%%%%%%%%%%%%%%%%%%%%%%%%%%%%%%%%%%%%%%%%%%%%%%%%%%%%%%%%%%%%%
%                                                                                                                                                   %
%                                                                                                                                                   %
%             Summary                                                                                                                          %
%                                                                                                                                                   %
%                                                                                                                                                   %
%%%%%%%%%%%%%%%%%%%%%%%%%%%%%%%%%%%%%%%%%%%%%%%%%%%%%%%%%%%%%%%%%%%%%%%%%%%%%%%%%%%%%%%%%%%%%%%%%%%%

\section{Summary}

Lattice gauge theory is a promising framework to reveal the gauge/gravity duality and the quantum effects of gravity from the gauge theory side.  
In this paper,  we have investigated the duality from the lattice simulations of 
supersymmetric $SU(N)$ Yang-Mills theory in 1+0 dimension with sixteen supercharges.
 The numerical results of the SUSY WTI have shown that 
  the Sugino lattice action that we used reproduces the correct continuum theory in the continuum limit. 
We have also estimated the internal energy of the black hole and found that it tended to approach the prediction of the gravity side at lower temperature, and the obtained power of the NLO term indicated the validity of the duality.

%The internal energy of the black hole have been estimated for sufficiently large $N$. 
At low temperature with small $N$, the instability related to the flat directions was observed.  
Whenever we encountered the instability, we changed $N$ to larger ones: $N = 14$ for $T_{{\rm eff}} = 0.5 \sim 5.0$ and $N= 32$ for $T_{{\rm eff}} = 0.375 \sim 0.5$, to avoid the problem. 
Also, all of the results for the internal energy of the black hole are ones obtained within the phase-quenched prescription at the fixed lattice sizes:
$L=16$ for $T_{\rm eff} < 1$ and $L=8$ for $T_{\rm eff} \ge 1$. 
Since the lattice spacing $a T_{\rm eff}=1/L$ was also fixed for each temperature, 
we have to take the continuum limit of the results for the NLO term.
For this purpose, the lattice simulations at different lattice spacings are ongoing.

\section*{Acknowledgment}
The numerical calculations were performed by using the RIKEN Integrated Cluster of Clusters (RICC) facility,
RIKEN's K computer and KEK supercomputer.
D.K. thank RICC's system engineers for hospitality, as well as staff of HPCI helpdesk (KEI computer).

\end{document}